# Toward understanding the $S_2$-$S_3$ transition in the Kok Cycle of Photosystem II: Lessons from Sr-Substituted Structure

Muhamed Amin[1,2,3]*, Divya Kaur[4,5,6], M.R. Gunner[5,6], Gary Brudvig[7]

[1] Rijksuniversiteit Groningen Biomolecular Sciences and Biotechnology Institute, University of Groningen, Groningen, Netherlands

[2] Physics Department, University College Groningen, University of Groningen, Groningen, Netherlands

[3] Centre for Theoretical Physics, The British University in Egypt, Sherouk City 11837, Cairo, Egypt

[4] Department of Physics and Department of Chemistry, Brock University, 500 Glenridge Avenue, St. Catharines, ON Canada L2S 3A1

[5] Department of Chemistry, The Graduate Center of the City University of New York, New York, NY 10016, United States

[6] Department of Physics, City College of New York, New York, NY 10031, United States

[7] Department of Chemistry, Yale University, New Haven, Connecticut 06520-8107, United States

m.a.a.amin@rug.nl

**ABSTRACT:** Understanding the water oxidation mechanism in Photosystem II (PSII) stimulates the design of biomimetic artificial systems that can convert solar energy into hydrogen fuel efficiently. The $Sr^{2+}$ substituted PSII is active but slower than with the native $Ca^{2+}$ as an oxygen evolving catalyst. Here, we use Density Functional Theory (DFT) to compare the energetics of the $S_2$ to $S_3$ transition in the $Mn_4O_5Ca^{2+}$ and $Mn_4O_5Sr^{2+}$ clusters. The calculations show that deprotonation of the water bound to $Ca^{2+}$ (W3), required for the $S_2$ to $S_3$ transition, is energetically more favorable in $Mn_4O_5Ca^{2+}$ than $Mn_4O_5Sr^{2+}$. In addition, we have calculated the $pK_a$ of the water that bridges Mn4 and the $Ca^{2+}/Sr^{2+}$ in the $S_2$ using continuum electrostatics. The calculations show that the $pK_a$ is higher by 4 pH units in the $Mn_4O_5Sr^{2+}$.

The oxygen evolving complex (OEC) is a unique natural bioinorganic cluster that catalyzes the water oxidation reaction in the 5-steps ($S_0$, $S_1$, $S_2$, $S_3$, $S_4$) Kok cycle.[1,2] The core of the OEC contains a metal cluster of four Mn and one $Ca^{2+}$ connected through bridging oxygens.[2–4] $Ca^{2+}$ depletion[5,6] blocks the $S_2$-$S_3$ transition, while replacing $Ca^{2+}$ with $Sr^{2+}$ reduces the catalytic activity.[7–10]

Calcium and strontium belong to group 2 alkaline earth metals in the periodic table. Thus, they are chemically similar and have a stable oxidation state of +2. However, $Ca^{2+}$ is a stronger Lewis acid, which indicates that aqua-$Ca^{2+}$ compounds have a lower $pK_a$ than aqua-$Sr^{2+}$ (measured $pK_a$ is 2 pH unit lower). This difference in proton affinity of the bound waters may be the reason for the difference in the catalytic activity in the Sr-substituted PSII.[10–12] Here, we use Density Functional Theory (DFT) to compare the energetics of the $S_2$-$S_3$ transition in the native and Sr-substituted PSII.

Experimental[13] and theoretical studies[14–16] have proposed that the $S_2$-$S_3$ transition passes through an intermediate step in which the $S_2$ EPR signal changes from the multiline g=2 signal to the g=4.1. In the g=4.1 EPR state Mn1, Mn2, Mn3 are in the IV oxidation state, while Mn4 is in the III state (Fig 1).[17] In the g=2 redox isomer M1 is $Mn^{3+}$ while M4 is $Mn^{4+}$. In addition, time-resolved photothermal beam deflection measurements suggest that a proton is released from the OEC or surroundings when the nearby Tyr, $Y_z$, is oxidized before Mn oxidation in the $S_2$-$S_3$ transition.[18,19]

Based on classical electrostatic calculations and DFT study[14], we previously proposed that the $S_2$-$S_3$ transition starts by the transition from g=2 to g=4.1 structure followed by deprotonation of the W3 $Ca^{2+}$ ligand.[20] This is coupled to the protonation of HIS190 upon the oxidation of the secondary donor $Y_z^*$. The deprotonated W3 moves toward Mn4 adding the sixth ligand to its coordination shell to facilitate its oxidation to IV state. Similar mechanisms have been proposed by previous theoretical[21–24] and experimental[10] studies.

Here, we compare the energies of two structures of the $S_2$ g=4.1 state, **A** in which HIS190 and W3 are neutral (Figure 1A) and **B** with protonated HIS190$^+$ and W3 is a OH$^-$ bridge between Mn4 and $Ca^{2+}$ (Figure 1B) in both $Mn_4O_5Ca^{2+}$ and $Mn_4O_5Sr^{2+}$ clusters. The structures were optimized at the DFT level using the B3LYP functional and 6-31G(d) basis sets for N, O, C and H atoms, while SDD are used for Mn, Ca and Sr. All the Mn ions are in the high spin state. Furthermore, the energies are compared using different levels of theory; B3LYP/6-31G+(d) and B97D/6-31G+(d).[25,26]

The energy differences between the **A** and **B** states ($\Delta G_{(B-A)}$) at different level of theory are shown in (Table 1). In general, the **B** state (protonated HIS190 and hydroxyl on W3) is always more favorable for the $Mn_4O_5Ca^{2+}$ than the $Mn_4O_5Sr^{2+}$ cluster. The large energy difference obtained for the $Mn_4O_5Sr^{2+}$ cluster using B3LYP/6-31G(d) level of theory indicates the importance of including diffuse functions in the basis sets when modeling large ions. These diffuse functions provide flexible representation to the tail part of the atomic orbitals further from the nucleus. [25,26]

$Sr^{2+}$ is larger than $Ca^{2+}$ by 0.1Å, which elongates the interatomic distances between the $Sr^{2+}$ and the rest of atoms in the Mn cluster. This is seen in the optimized structures of the **A** and **B** states with $Ca^{2+}$ and $Sr^{2+}$ clusters (Table 2). In addition, the dispersion interaction between the metal and the water ligand is expected to push the water away in case of $Sr^{2+}$, which will result in smaller electrostatic interactions and a higher $pK_a$. This is found for aqua-$Ca^{2+}$ and aqua-$Sr^{2+}$ compounds, where the water bound to $Sr^{2+}$ have a higher $pK_a$ than those bound to $Ca^{2+}$. Thus, the $Sr^{2+}$ structure is more stable with neutral W3 (Figure 1**A**). However, with $Ca^{2+}$, W3 deprotonates forming a hydroxide that moves to bridge Mn4 and $Ca^{2+}$ (Figure 1**B**).

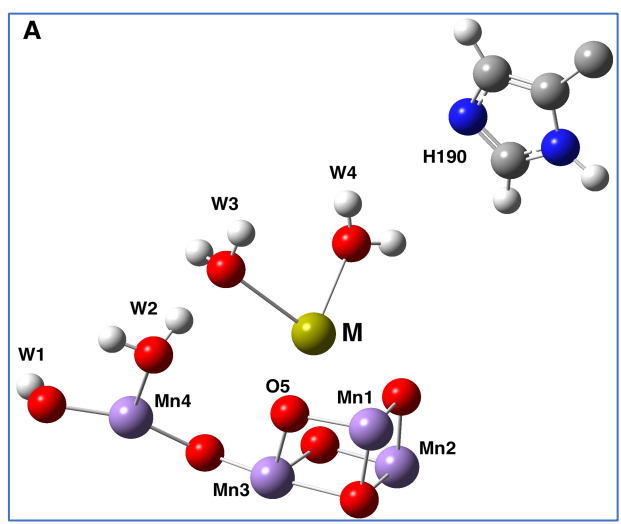

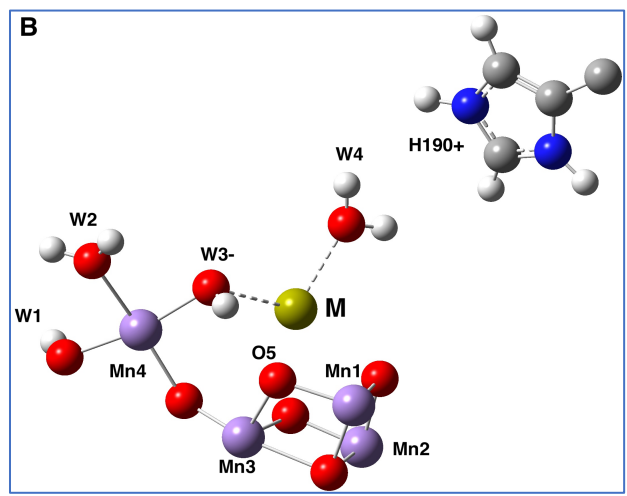

**Figure 1**. **A** represents the $S_2$ state with HIS190 neutral. **B** represents the $S_2$ state with $HIS190^+$ protonated. **M** is $Ca^{2+}$ or $Sr^{2+}$. Mn1, Mn2, Mn3 are in the IV oxidation state. Mn4 is III. The rest of atoms in the model were removed for clarity.

The optimized DFT structures show that Mn-$Sr^{2+}$ distances are in general longer than Mn-$Ca^{2+}$. In the **A** state the $Sr^{2+}$-W3(HOH) distance is longer by 0.1Å than $Ca^{2+}$-W3(HOH), while in the **B** state the $Sr^{2+}$-W3(OH)⁻ distance is 0.2Å longer. The Mn1 to $Ca^{2+}$, is longer because $Ca^{2+}$ moves significantly toward Mn4 after the deprotonation of W3.

**Table 1**. The $\Delta G_{(B-A)}$ DFT energies.

|  | B3LYP/ 6-31G(d) | B3LYP/6-31G+(d) | B97D/6-31G+(d) |
|---|---|---|---|
| $Mn_4O_5Ca^{2+}$ | -2.3 | -8.7 | -6.8 |
| $Mn_4O_5Sr^{2+}$ | 13.0 | -8.0 | -6.4 |

Energy differences are expressed in Kcal/mol. The transition from A to B state is more favorable in the $Mn_4O_5Ca^{2+}$ cluster than the $Mn_4O_5Sr^{2+}$

To further compare the $Mn_4O_5Ca^{2+}$ the $Mn_4O_5Sr^{2+}$, we calculated the $pK_a$ of W3 in the **A** state for both clusters using continuum electrostatics.[27,28] W3 has a $pK_a$ of 6.5 for the $Mn_4O_5Ca^{2+}$ and 10.3 in the $Mn_4O_5Sr^{2+}$. The lower pK of W3 in $Ca^{2+}$ structure is expected as W3 is significantly closer to the positively charged ions ($Ca^{2+}$, Mn4(III) and Mn1(IV) (Table 2). This conclusion is supported by the DFT calculations, which show that **B** is lower energy than **A** indicating the easier deprotonation in case of the $Mn_4O_5Ca^{2+}$.

An open question is what the source of the proton which is released after $Y_z$ is oxidized but before the OEC

advances to the $S_3$ state.[3,29] As there are no protons bound to the bridging oxygens in the S2 state, the donors are likely to be terminal water ligands bound to Mn4 water[30,31] or to Ca.[32] Previous studies have shown the Mn4-bound water W1 to be upon the formation of tyrosyl radical, however the proton is trapped by the neaby acceptor D61.[16,33,34]

The present study utilizes the $S_2$ g=4.1 models for $Ca^{2+}$ and $Sr^{2+}$ to understand the nature of deprotonation event. Our DFT calculations support the deprotonation of W3 in the $S_2$ to $S_3$ transition, which is also supported by the XFEL structures comparing $S_1$, $S_2$ and $S_3$ states.[28]

Table 2. Interatomic distances in A and B states.

|  | A | | B | |
| --- | --- | --- | --- | --- |
|  | $Ca^{2+}$ | $Sr^{2+}$ | $Ca^{2+}$ | $Sr^{2+}$ |
| Mn1 | 2.50 | 3.58 | 3.49 | 3.67 |
| Mn4 | 4.35 | 4.51 | 3.58 | 2.64 |
| W3 | 2.55 | 2.66 | 2.39 | 2.58 |
| W4 | 2.33 | 2.50 | 2.35 | 2.53 |
| O5 | 2.66 | 2.78 | 2.66 | 2.77 |

All distances are reported in Å. In general, the interatomic distances are longer for $Sr^{2+}$.

## Supporting Information

The Supporting Information includes details of the continuum electrostatic calculations and the optimized atomic coordinates of the optimize **A** and **B** structures.

## Acknowledgements

The authors acknowledge computational resources from the support by the U.S. Department of Energy, Office of Science, Office of Basic Energy Sciences, Division of Chemical Sciences, Geosciences, and Biosciences via Grants DESC0001423 (M.R.G. and V.S.B.), DE-FG02-05ER15646 (G.W.B.).